\documentclass{moriond}

\def\be{\begin{equation}}
\def\ee{\end{equation}}
\def\bea{\begin{eqnarray}}
\def\eea{\end{eqnarray}}

\def\PW{{\rm{W}}}
\def\PZ{{\rm{Z}}}
\def\Pe{{\rm{e}}}
\def\Pj{{\rm{j}}}
\def\rL{{\rm{L}}}
\def\rT{{\rm{T}}}

\newcommand{\Photo}{\includegraphics[height=35mm]{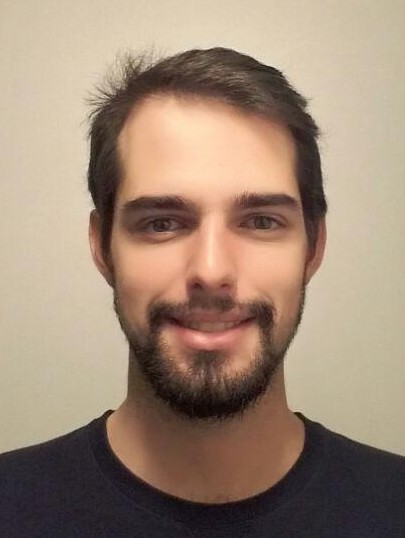}}

\begin{document}
\vspace*{4cm}
\title{Precise predictions for polarised weak bosons at the LHC}

\author{Giovanni Pelliccioli}

\address{Max-Planck-Institut f\"ur Physik, F\"ohringer Ring 6, 80805 M\"unchen, Germany}

\maketitle\abstracts{Precise and accurate Standard Model predictions are needed for polarised weak bosons in LHC processes, in order to perform template fits of data and to enhance the sensitivity to possible new-physics effects. We have proposed a general strategy to compute polarised cross-sections including radiative QCD and electroweak corrections to the production and decay of bosons. The method relies on the pole approximation and the separation of polarisation states at amplitude level. After showing some details of the theoretical definition, we present results relevant for LHC di-boson polarisation analyses in semi-leptonic decay channels.}

\section{Motivation}
The LHC luminosities accumulated in Run 2 and foreseen in next runs
(Run 3, High-Lumi) will enable precise measurements of electroweak (EW) processes with
one or more weak bosons ($\PW^\pm,\,\PZ$).
Accessing the polarisation state of weak bosons represents a crucial step towards a better understanding
of the electroweak-symmetry-breaking mechanism realised in nature, providing us with important
probes of the Standard Model (SM) gauge and Higgs sectors, as well as high discrimination power between
SM and beyond-the-SM (BSM) dynamics.
Unfortunately, extracting EW-boson polarisation states is hampered by the unstable nature of
massive gauge bosons, making it impossible to directly detect polarised bosons. However, the polarisation
state of $\PW$ and $\PZ$ bosons leaves trace in the kinematic distributions of the decay products.
The typical approach used in Run-1 polarisation analyses relied on the
extraction of coefficients from decay-product angular distributions \cite{ATLASCMSVJATLASCMSTOP}.
More recently, a template-fit approach was introduced by ATLAS and CMS in the analysis programme,
leading to polarisation measurements in di-boson inclusive production \cite{ATLASCMSWZ}
and scattering \cite{CMS:2020etf}.
Sensitivity studies for polarisation measurements in the High-Lumi and High-Energy stages of the LHC are promising
\cite{EXPSENS}.
In order to enable fits of LHC data with polarised templates, we need three ingredients from the theory side:
(1) proper control on the definition of polarised signals,
(2) high accuracy in the perturbative and non-perturbative predictions,
and (3) clever ideas to enhance the sensitivity to polarisation in the LHC environment.

\section{Theoretical definition of polarised signals}
A natural definition of polarised signals can be identified in the case of processes
that are described by resonant diagrams (in a certain on-shell approximation, with a given gauge choice).
\begin{figure}
  \centering
  \includegraphics[width=0.35\linewidth]{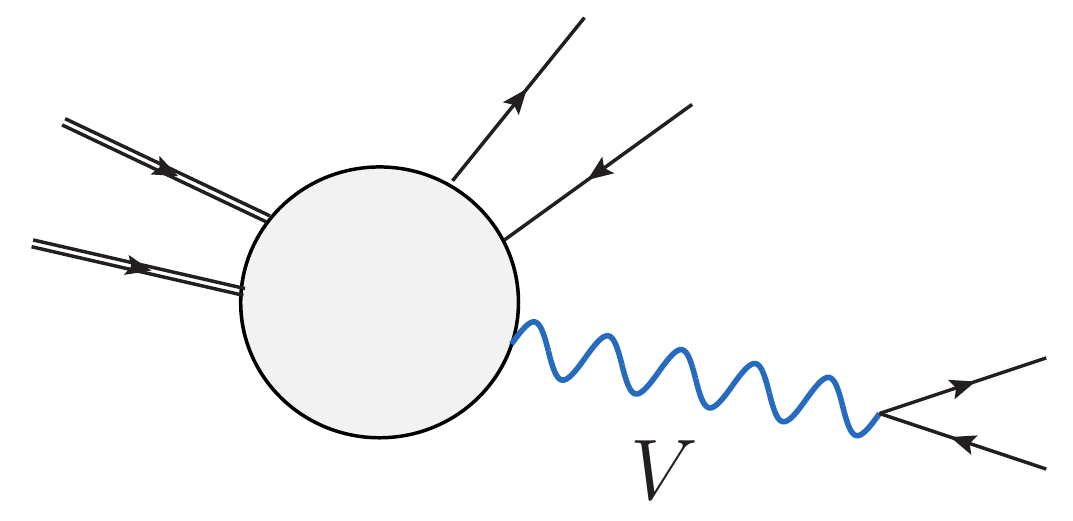}
  \hfill
  \caption[]{Generic resonant contribution to single-boson production and decay.}
  \label{fig:Vres}
\end{figure}
In the t'Hooft-Feynman gauge, the (unpolarised) amplitude for resonant diagrams as the one in Fig.~\ref{fig:Vres} can be written as,
\be
\mathcal{A}^{\rm unp} = {\mathcal{P}}_{\mu}\,\,{\frac{-g^{\mu\nu}
  }{k^2-M_V^2+iM_V\Gamma_V}}\,\,{\mathcal{D}}_{\nu} = {\mathcal{P}}_{\mu}\,\,{\frac{\sum_{\lambda'}\varepsilon_{\lambda'}^{\mu}\varepsilon_{\lambda'}^{*\nu}}{k^2-M_V^2+iM_V\Gamma_V}}\,\,{\mathcal{D}}_{\nu}\,.
\ee
The sum runs over physical polarisation modes as unphysical modes are canceled by Goldstone-boson contributions. Replacing the sum with a single term we obtain a {polarised amplitude},
\be
\mathcal{A}_\lambda =  {\mathcal{P}}_{\mu}\,\,{\frac{\varepsilon_{\lambda}^{\mu}\varepsilon_{\lambda}^{*\nu}}{k^2-M_V^2+iM_V\Gamma_V}}\,\,{\mathcal{D}}_{\nu}\,, \quad \textrm{for polarisation $\lambda=\rL,+,-$}.
\ee
Therefore, the squared unpolarised amplitude is written an incoherent sum of squared polarised amplitudes plus interference terms:
\be\label{eq:unpolTOpol}
         {|\mathcal{A^{\rm unpol}}|^2} =
           \sum_{\lambda}\left|
           \mathcal{A}_{\lambda}\right|^2
         +
           \sum_{\lambda \neq \lambda'}
           \mathcal{A}_{\lambda}^{ *}\mathcal{A}_{\lambda'}
\ee
Up to flux, symmetry and phase-space factors, we obtain a natural definition of \emph{polarised cross section}
proportional to $\left|\mathcal{A}_{\lambda}\right|^2$ \cite{Ballestrero:2017bxn}.
Note that polarisation vectors (and therefore polarised signals) are defined in a specific Lorentz frame.
As mentioned above the decay-product distributions reflect the polarisation state of the decayed boson,
thus the polarisation fractions ($f_{\rL}, f_{\pm}$) can be extracted from the unpolarised
decay-angle distribution \cite{BernStirling},
\be
\frac{\rm d\sigma}{{\rm d}\!\cos\theta^*}\frac1{\sigma}= \frac{3f_{\rL}}4 (1-\cos^2{\theta^*})+  \frac{3f_{+}}8 (1+\cos^2{\theta^*}+2c_{\rm RL}\cos{\theta^*}) + \frac{3f_{-}}8(1+\cos^2{\theta^*}-2c_{\rm RL}\cos{\theta^*})\,,
\label{eq:mastereq}
\ee
by means of projections onto suitable spherical harmonics.
However, this strategy is not viable with more than one bosons, with radiative corrections modifying the decay structure,
and in the presence of cuts on decay products \cite{BernStirling,Ballestrero:2017bxn}.
On the contrary, the polarised-signal definition of Eq.~\ref{eq:unpolTOpol} can be systematically applied
in the presence of more bosons and including such realistic effects.

Since in general both resonant and non resonant diagrams contribute to
any LHC process already at leading order (LO), only resonant diagrams
must be selected, recovering the EW gauge invariance by means of a narrow-width
or pole approximation \cite{NWADPA}.
Then, separating polarised amplitudes is straightforward. This strategy
can be applied to processes with any number of EW bosons
and a general method was proposed \cite{Denner:20202021,Le:2022lrp}
to extend it also to the presence of next-to-leading-order
(NLO) QCD and EW corrections (both to production and to decay).
This has been successfully applied to all di-boson processes in the fully leptonic decay
channel up to NNLO QCD and NLO EW accuracy
\cite{Denner:20202021,Poncelet:2021jmj,Le:2022lrp}.
The extension of the method to the NLO matching to parton showers is ongoing 
in the {\scshape{Powheg-Box-Res}} framework \cite{PZ23}.

\section{Polarised di-boson production with semi-leptonic decays}
Very recently, the first calculation of di-boson inclusive production in the semi-leptonic decay channel
has been carried out at NLO QCD in the presence of polarised bosons \cite{Denner:2022riz}.
This was done in the {\scshape{MoCaNLO}} Monte Carlo code with 
tree-level and one-loop SM amplitudes calculated with {\scshape{Recola}} \cite{Actis:2016mpe}
and {\scshape{Collier}}\cite{Denner:2016kdg} libraries. 
The considered process is ${ \rm p \, p \rightarrow \PZ (\rightarrow \rm e^+\, e^-) \, \PW^+ (\rightarrow \rm jets)}$
 at $\sqrt{s}=13.6$TeV,
with fiducial selections that mimic those of a recent CMS analysis \cite{CMS:2021xor}. Two setups are studied, depending on the two-light-jet (resolved) or
one-fat-jet (unresolved) hadronic decay of the $\PW$ boson. A boosted regime for
both bosons is considered ($p_{{\sf T}, V}>200$GeV). Polarisation states
(L=longitudinal, T=transverse) are defined in the {centre-of-mass frame} of the di-boson system.
The integrated cross sections are shown in Table~\ref{tab:semilept}.
\begin{table}[t]
  \caption[]{Fiducial cross sections and fractions for (un)polarised $\PZ\PW^+$ production with semi-leptonic decays
for the two setups. Monte Carlo errors (in parentheses) and QCD-scale uncertainties (in percentage) are shown.}
  \label{tab:semilept}
  \vspace{0.4cm}
  \begin{center}
    \scriptsize
    \begin{tabular}{ccccccc}%
      \hline %
      state  &  $\sigma_{\rm LO}$ [fb]  & $f_{\rm LO}[\%]$&  $\sigma_{\rm NLO}$ [fb] & f$_{\rm NLO}[\%]$ & {$K_{\rm NLO}$} & {$K_{\rm NLO}^{\rm (no\,g)}$}      \\
      \hline
      \multicolumn{7}{l} {resolved setup, $\PZ(\Pe^+\Pe^-)\PW^+(\Pj\Pj)$}\\
      \hline
      unpol. & $  1.8567  ( 2 )^{+ 1.2 \%}_{- 1.4 \%}$ &$100$&        $ 3.036(2)  ^{+ 6.8 \%}_{- 5.3 \%} $ &$ 100 $ & { $         1.635 $}  & $ 1.033$\\
      $\PZ^{\,}_{\rL}\PW^{+}_{\rL}$& $  0.64603  ( 5 ) ^{+ 0.2 \%}_{- 0.6 \%}$  &{ $ 34.8$}&      $ 0.6127  ( 4 ) ^{+ 0.9 \%}_{- 0.7 \%} $  &{ $ 20.2 $} & $        0.948 $  & $1.031 $\\
      $\PZ^{\,}_{\rL}\PW^{+}_{\rT}$& { $  0.08687  ( 1 ) ^{+ 0.2 \%}_{- 0.6 \%}$}  &$ 4.7$&       { $ 0.17012  ( 6 ) ^{+ 8.6 \%}_{- 6.8 \%} $}  & $ 5.6 $& { $          1.958 $}  & $ 0.967 $\\
      $\PZ^{\,}_{\rT}\PW^{+}_{\rL}$& { $  0.08710  ( 1 ) ^{+ 0.1 \%}_{- 0.6 \%}$}  &$ 4.7$&       { $ 0.24307  ( 7 ) ^{+ 10.2 \%}_{- 8.2 \%} $}  &$ 8.0 $ & { $         2.791  $} & $1.017 $ \\
      $\PZ^{\,}_{\rT}\PW^{+}_{\rT}$& { $  0.97678  ( 7 ) ^{+ 2.0 \%}_{- 2.2 \%}$}  &$ 52.6$&      { $ 2.0008  ( 7 ) ^{+ 8.9 \%}_{- 7.1 \%} $}  &$ 65.8 $ & { $        2.048 $}  & $1.059 $\\
      { interf. }& { $ 0.0595(1) $  }&{ $3.2 $}&      { $ 0.009(2)$}  &{ $0.4 $} & $-$ & $- $\\
      \hline
      \multicolumn{7}{l}{  unresolved setup, $\PZ(\Pe^+\Pe^-)\PW^+({\rm J})$} \\
      \hline
      unpol. &$  1.6879  ( 2 )^{+ 1.9 \%}_{- 2.1 \%}$  &$100$&        $ 3.112(2)  ^{+ 7.6 \%}_{- 6.1 \%} $ &$100 $  & { $         1.843 $} & $ 1.193$ \\
      $\PZ^{\,}_{\rL}\PW^{+}_{\rL}$& $  0.61653  ( 5 ) ^{+ 1.0 \%}_{- 1.3 \%}$  &{ $36.5 $}&      $ 0.6799  ( 5 ) ^{+ 0.9 \%}_{- 0.7 \%} $ & { $21.9 $} & $         1.103 $ & $1.170$ \\
      $\PZ^{\,}_{\rL}\PW^{+}_{\rT}$& { $  0.06444  ( 1 ) ^{+ 0.7 \%}_{- 1.0 \%}$}  &$3.8 $&       { $ 0.17584  ( 6 ) ^{+ 10.8 \%}_{- 8.6 \%} $} & $5.7 $ & { $          2.729  $}& $1.158 $  \\
      $\PZ^{\,}_{\rT}\PW^{+}_{\rL}$& { $  0.07437  ( 1 ) ^{+ 0.6 \%}_{- 0.9 \%}$}  &$4.4 $&       { $ 0.24742  ( 8 ) ^{+ 11.0 \%}_{- 8.9 \%} $} & $8.0 $ & { $          3.327  $}& $ 1.193 $  \\
      $\PZ^{\,}_{\rT}\PW^{+}_{\rT}$& { $  0.88233  ( 9 ) ^{+ 2.9 \%}_{- 2.9 \%}$}  &$52.3 $&      { $ 2.0041  ( 8 ) ^{+ 9.6 \%}_{- 7.7 \%} $} & $64.3 $ & { $         2.271 $} & $ 1.227 $ \\
      { interf.} & { $ 0.0503(3) $}  &{ $3.0 $}&      { $0.004(2) $}  &{ $0.1 $} & $-$ & $- $\\
      \hline
    \end{tabular}
  \end{center}
\end{table}
The large NLO QCD corrections mostly come from gluon-initiated real contributions.
A large LL fraction characterises the boosted regime, compared to inclusive setups. The longitudinal state is unsuppressed, due
to the presence of the triple-gauge coupling. 
Sizeable differences are found between the two setups at LO, owing to jet recombination, while they become smaller at NLO.
Interference contributions are very small, especially at NLO (less than 0.5\% of the total).

\begin{figure}
  \begin{minipage}{0.5\linewidth}
    \centerline{\includegraphics[width=0.8\linewidth]{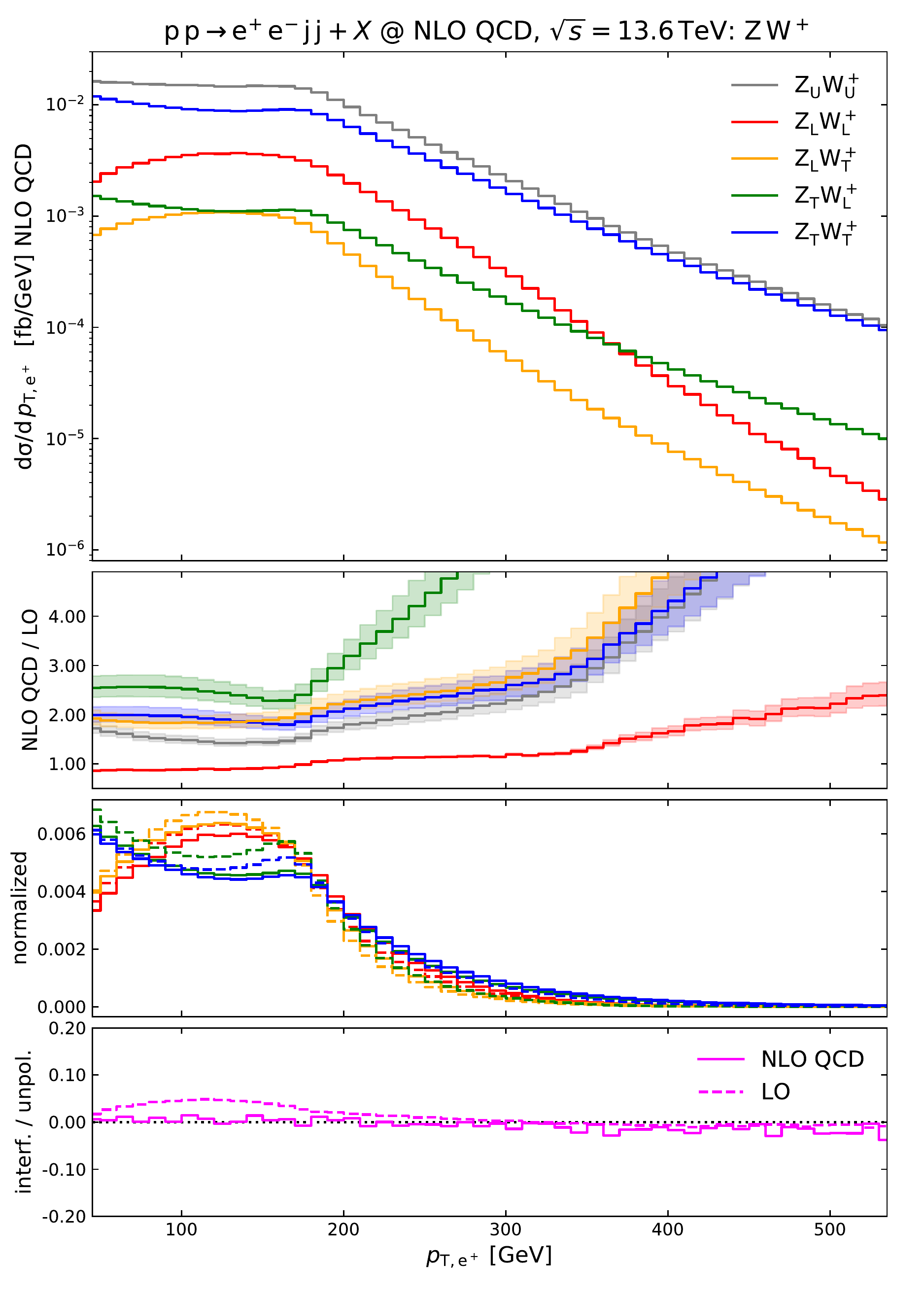}}
  \end{minipage}
  \hfill
  \begin{minipage}{0.5\linewidth}
    \centerline{\includegraphics[width=0.8\linewidth]{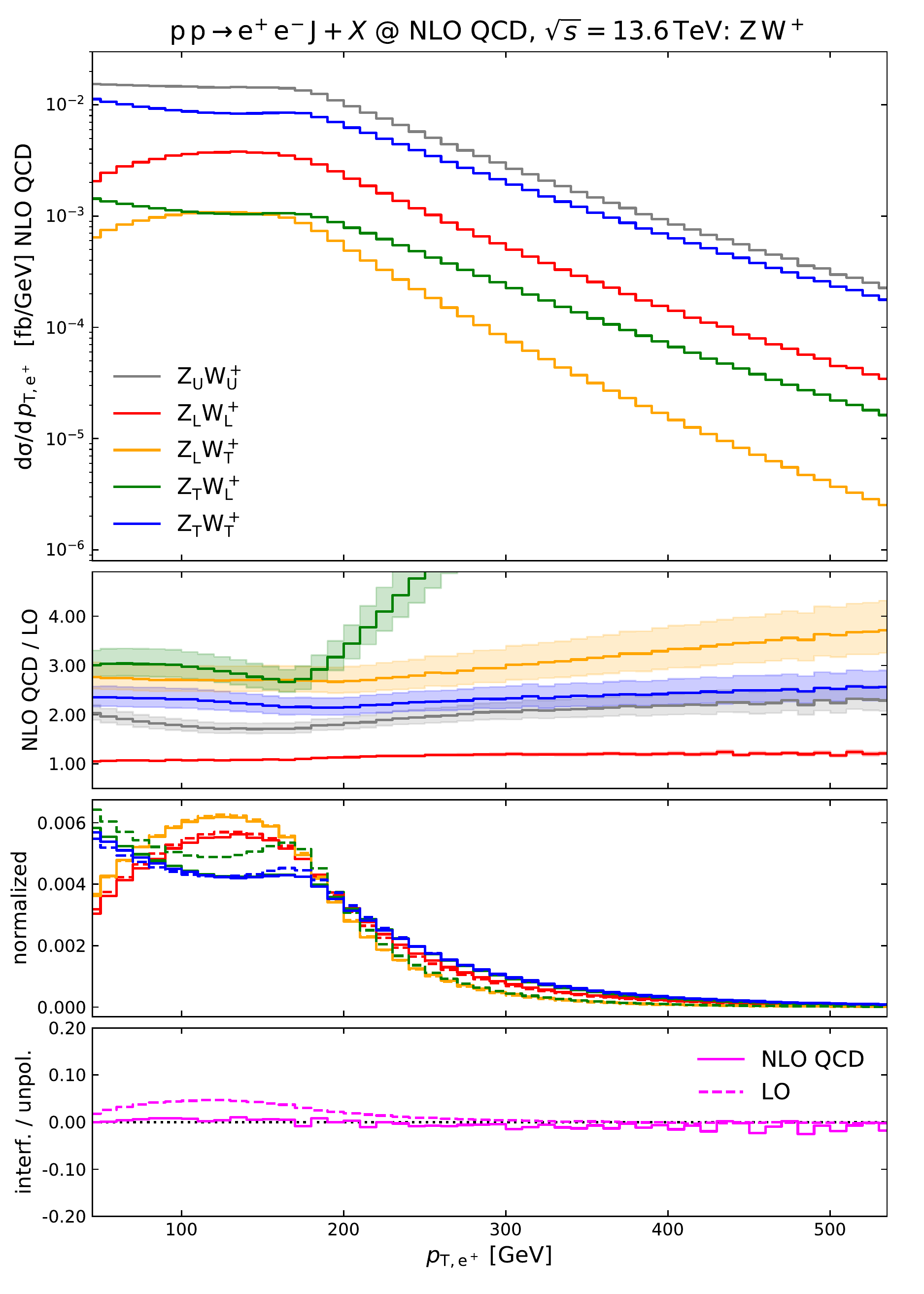}}
  \end{minipage}
  \hfill
  \caption[]{
   Distributions in the positron transverse momentum for polarised and unpolarised $\PZ\PW^+$ production in the semi-leptonic decay channel,
   for the resolved (left) and unresolved (right) setups.
   From top down:
   differential cross sections,
   NLO QCD $K$-factors,
   normalised distributions (unit integral),
   relative interference contribution.}
  \label{fig:ptep}
\end{figure}

It is essential to study the differential distributions for (doubly) polarised signals,
which typically feature an enhanced discrimination power between longitudinal and transverse states. In Fig.~\ref{fig:ptep}
the transverse momentum of the positron is considered for the two setups and for the various polarisation states.
There is a clear sensitivity to the $\PZ$-boson polarisation in the low-transverse-momentum range. The faster decrease
of the curves in the high-$p_{\rm T,e^+}$ regime at LO in the resolved setup gives larger $K$-factors than in the unresolved setup.
A number of other observables have very strong discrimination power
amongst the doubly polarised states \cite{Denner:2022riz}.

\section{Conclusions}
The extraction of polarised-boson signals is of high interest for the LHC community, with ongoing (and upcoming)
data analyses triggering new phenomenological studies and theoretical developments.
So far most of the theoretical effort has been devoted to SM fixed-order predictions (QCD and EW corrections), to the automation in Monte Carlo codes, and to the
search for polarisation-sensitive observables, with special focus on di-boson production.
The calculation of polarised predictions matched to parton-shower is ongoing. In the coming years, it will be desirable to achieve accurate SM predictions
for polarised vector-boson scattering (the \emph{golden channel} for polarisation) and to study
new-physics effects in the production and the decay of polarised bosons.

\end{document}